\documentclass[%
 reprint,
 amsmath,amssymb,
 aps,
]{revtex4-1}

\usepackage{graphicx}
\usepackage{dcolumn}
\usepackage{bm}
\usepackage{hyperref}
\usepackage{color}

\begin{document}

\preprint{APS/123-QED}

\title{Orbital Angular Momentum Generation by a Point Defect in Photonic Crystals}

\author{Menglin L. N. Chen}
\affiliation{%
	Electromagnetics and Optics Lab, Department of Electrical and Electronic Engineering, The University of Hong Kong, Hong Kong
}%
\author{Wei E. I. Sha}%
\email{weisha@zju.edu.cn}
\affiliation{%
The Innovative Institute of Electromagnetic Information and Electronic Integration,
College of Information Science and Electronic Engineering, Zhejiang University, Hangzhou 310027, P. R. China.
}%
\author{Li Jun Jiang}
\email{jianglj@hku.hk}
\affiliation{%
Electromagnetics and Optics Lab, Department of Electrical and Electronic Engineering, The University of Hong Kong, Hong Kong
}%


\begin{abstract}
As an attractive degree of freedom in electromagnetic (EM) waves, the orbital angular momentum (OAM) enables infinite communication channels for both classical and quantum communications. The exploration of OAM generation inspires various designs involving spiral phase plates, antenna arrays, metasurfaces, and computer-generated holograms. In this work, we theoretically and experimentally demonstrate an approach to producing OAM carrying EM waves by a point defect in three-dimensional (3D) photonic crystals (PCs). Simultaneous excitation of two vibrational-defect states with an elaborately engineered phase retardation generates a rotational state carrying OAM. Through converting guided waves in a line defect to localized waves in a point defect and then to radiated vortex waves in free space, the lowest four OAM-mode emitters, i.e., OAM indices of $\pm1$ and $\pm2$, are successfully realized. This work offers a physical mechanism to generate OAM by PCs, especially when the OAM generation is to be integrated with other designs.
\end{abstract}

\maketitle


\section{\label{sec:level1}Introduction}

Electromagnetic (EM) waves carry both linear and angular momenta. Recently, the angular momentum has been attracting much attention due to the extra degrees of freedom introduced into EM waves. It is well known that circularly polarized waves carry spin angular momentum (SAM), which characterizes the spin feature of a photon. Unlike the SAM, orbital angular momentum (OAM) manifests the orbital rotation of photon. It describes structured waves possessing a helical wave front. The number of twists in the helical wave front identifies each OAM state. Therefore, OAM-carrying waves cause distinctive phase structures and allow for promising applications, ranging from classical to quantum regimes~\cite{radio,wang2012terabit,gibson2004free,mair2001entanglement,2007rev,2016OAMexperiment,qiuchengwei,liukang}.

Since OAM is associated with the spatially variant phase factor, $e^{il\phi}$ (where $\phi$ is the azimuthal angle and $l$ is the OAM index)~\cite{allen}, a great many designs that manipulate the spatial phase have been proposed for OAM generation, such as spiral phase plates (SPPs)~\cite{SSP_SREP,SPP_analytical,SPP_2005}, metasurfaces~\cite{mohammadi2010orbital,capasso,shuangzhang,2013Li,cuitiejun,menglin1}, computer-generated holograms~\cite{1990CGH, 1992CGH} and other novel prototypes~\cite{gumin}. Although Maxwell's equations govern the behavior of EM waves from microwave to optical frequencies, scatterers respond distinctively due to the dispersive nature of materials. Consequently, engineered designs adopting the same working principle face different challenges as the frequency evolves. For example, in optics, metasurfaces are promising candidates for OAM generation because of their ultrathin configuration and tuning flexibility~\cite{Capasso_review}. However, they usually suffer from low-efficiency transmission, caused by reflection, diffraction and ohmic loss~\cite{APL_spin_to_orbital}. At microwave frequencies, the conversion efficiency can be promisingly high, but scatterers of the metasurface need to be fully aligned and stacked to achieve a desired EM response~\cite{menglin2}. Meanwhile, in contrast to optical regime, where light manipulation can be conveniently actualized using versatile optical devices including polarizers, beam splitters, holograms, and diffraction gratings, the functions of some optical devices cannot be trivially replicated at microwave frequencies. Interestingly, among all the building blocks and devices, photonic crystals (PCs), can be universally and scalably applied in both the microwave and optical fields due to the scaling law.

The concept of PCs was proposed by Yablonovitch~\cite{Yablonovitch} and John~\cite{John} in $1987$. It is an analog of periodically arranged atoms and, therefore, opens up a band gap where no EM state is allowed to propagate. However, a localized state will be supported at the photonic band gap if a defect is introduced in PCs and the translation symmetry is broken~\cite{point_defect}. The defects, similar to dopants in semiconductors, are highly tunable by geometries and placements, which makes them useful in resonators, waveguides, filters, and couplers~\cite{pcbook1, pcbook2}.

\begin{figure}[htbp]
\centering
\includegraphics[width=0.8\columnwidth]{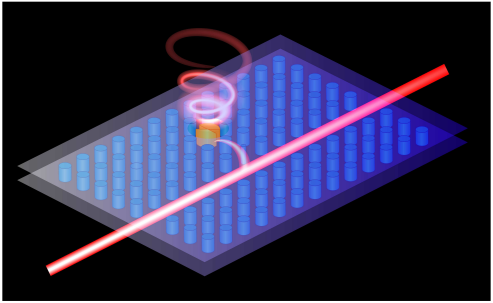}
\caption{A schematic representation of OAM generation in the proposed PCs. The guided waves in the line defect are coupled to the localized modes in the point defect and then radiate out through a circular opening. The OAM is introduced by the simultaneous excitation of two defect states with desired weights and phase difference.}
\label{cover}
\end{figure}

Most importantly, point-defect states are the whispering-gallery-like modes that can be assigned orbital angular momenta~\cite{new_twist}. In this paper, we theoretically and experimentally demonstrate an approach to producing OAM by superposing two point-defect states in three-dimensional (3D) PCs. The two defect states with orthogonal vibration modes are excited simultaneously in a point defect of PCs to emit a mixed rotational mode carrying OAM. Two PC structures are constructed to generate different OAM states with $l=\pm1$ and $l=\pm2$. In our designs, EM energy is transferred from guided waves in a line defect to localized resonating modes in a point defect and then to unbounded OAM states in free space, this being summarized in Fig.~\ref{cover}. Thus, the physical mechanism of OAM generation is distinguished from existing ones employed in metasurfaces and computer- generated holograms. As a proof of concept, all the theoretical and experimental investigations are to be implemented at the microwave regime.

\section{Methods and results}

\subsection{Defect modes in two-dimensional PCs}

\begin{figure}[htbp]
\centering
\includegraphics[width=\columnwidth]{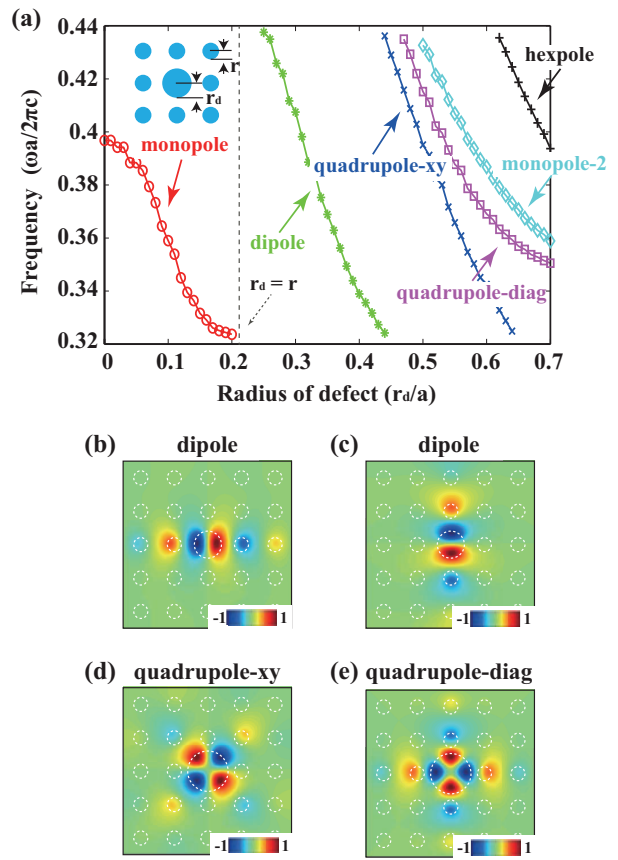}
\caption{The localized modes inside the photonic band gap introduced through a point defect. (a) The mode frequency as a function of radius $r_d$ of the defect rod. (b, c) The real parts of the electric field of the doubly degenerate dipole modes when $r_{d}=0.4a$. Their mode frequency is $0.3389c/a$. (d, e) The real parts of the electric field of two nondegenerate quadrupole modes when $r_{d}=0.6a$. Their mode frequencies are $0.3425c/a$ and $0.3691c/a$, respectively.}
\label{analytical}
\end{figure}

We calculate the band structure for transverse magnetic (TM) modes in a two-dimensional (2D) PC by using the finite-difference (FD) method (for details, see the Supplemental Material~\cite{SI}). The forbidden frequencies range from $0.3236c/a$ to $0.4382c/a$, where $a$ is the lattice constant of the PC and $c$ is the speed of light in vacuum. By introducing a point defect, localized modes will be formed within the band gap. In Fig.~\ref{analytical}, we show the frequencies of the defect modes as a function of radius $r_d$ of the defect rod (radius of bulk rods $r=0.2a$). The dielectric constant of the defect rod is set to be $8.5$, which is the same as that of bulk rods. In view of the fourfold ($C_4$) rotational symmetry, the dipole and hexapole defect modes are doubly degenerate, such that one can be reproduced from the other by rotating $90^\circ$. The quadrupole defect modes are nondegenerate and have two nodal planes. The quadrupole-xy mode has the nodal planes along the $x$ and $y$ directions, while they are along the diagonal directions for the quadrupole-diag mode.

\subsection{OAM generation from quadrupole defect states}

It is well known that a circularly polarized wave can be decomposed into two orthogonal linearly polarized waves with the same amplitude but a phase difference of $90^{\circ}$. Molecular mechanics also obeys the physics that two vibrational modes with a proper phase delay will generate a rotational mode~\cite{Landau}. Based on this concept, a rotational vortex beam carrying OAM can be generated by superposing two quadrupole modes as the two vibrational modes depicted in Figs.~\ref{analytical}(d,e).

We first mathematically model a linear superposition of the two modes,
\begin{equation}
E_z^{tot}(\bm{\rho}) = A E_z^1(\bm{\rho})e^{i\theta}+BE_z^2(\bm{\rho})
\label{Eq:1}
\end{equation}
where $E_z^1$ and $E_z^2$ are the normalized electric fields for the two quadrupole modes. $A$ and $B$ are the weights of two modes and $\theta$ is the relative phase of the mode $1$ with respect to the mode $2$.

\begin{figure}[htbp]
\centering
\includegraphics[width=\columnwidth]{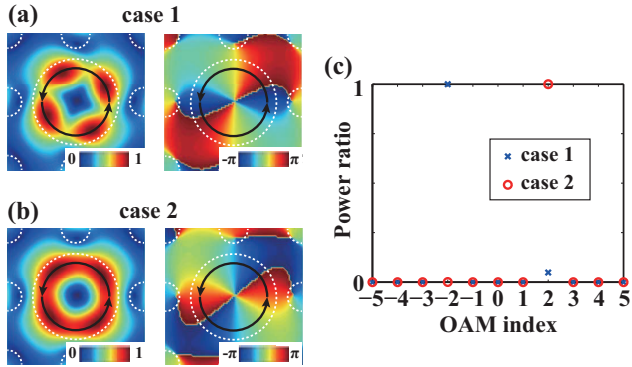}
\caption{Superposed electric-field patterns ($E_z$) of two quadrupole modes when $r_{d}=0.6a$. (a) The amplitude and phase patterns for case 1. (b) The amplitude and phase patterns for case 2. (c) The projection of the complex field along the azimuthal direction onto an orthonormal basis of $e^{il\phi}$ [denoted by the black circle in (a) and (b)]. Optimal weights and phase are chosen so that in each case, only one OAM order ($l=-2$ or $+2$) is dominant.}
\label{quad_analytical}
\end{figure}

The superposed electric field of the two quadrupole modes with $r_{d}=0.6a$ is investigated by varying the weights and relative phase. We project the superposed field onto an orthonormal basis of $e^{il\phi}$ along the azimuthal direction (denoted by the black circles in Fig.~\ref{quad_analytical}). As the weights and phase change, the projection results will be modified. A rotational mode with the spatial phase dependence of $e^{\pm i2\phi}$ is generated by the optimal weights ($A=B=1$) and phase ($\theta=\pm90^{\circ}$), as shown in Fig.~\ref{quad_analytical}. The phase profiles indicate an OAM index of $\pm2$ and a phase singularity can be clearly observed at the center of the amplitude pattern.

\begin{figure}[htbp]
\centering
\includegraphics[width=\columnwidth]{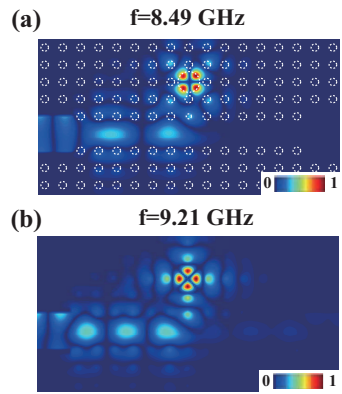}
\caption{Excited quadrupole states inside a 2D PC incorporating a small scatterer as a mode coupler. (a) The quadrupole-xy state. (b) The quadrupole-diag state.}
\label{quad_sim_2d}
\end{figure}

To mimic 2D PCs by 3D structures in software (the CST Microwave Studio), the array of finite-length dielectric rods is inserted between a lossless metallic parallel-plate waveguide. The height of the parallel-plate waveguide is sufficiently small to guarantee that only the transverse electromagnetic (TEM) mode propagates, where the electric field is uniform and only has the $z$ component. Hence, the supported TEM mode is compatible with the TM modes in 2D PCs. The quadrupole modes at the defect with a radius of $r_d=0.6a$ ($a=12$~mm) are excited through the guided wave in a line defect, as shown in Fig.~\ref{quad_sim_2d}. A normal rod that is adjacent to the line defect is replaced by a smaller rod (scatterer) with a radius of $r_s=0.1375a$ to enhance the coupling efficiency. The two eigenmodes are found to be at $f=8.49$~GHz and $f=9.21$~GHz. The states at the frequencies between $8.49$~GHz and $9.21$~GHz can be considered as the superposition of the two quadrupole modes with specific weights and relative phase [See Eq. (1)] according to the eigenmode expansion theory~\cite{chew}.

\begin{figure}[htbp]
	\centering
	\includegraphics[width=\columnwidth]{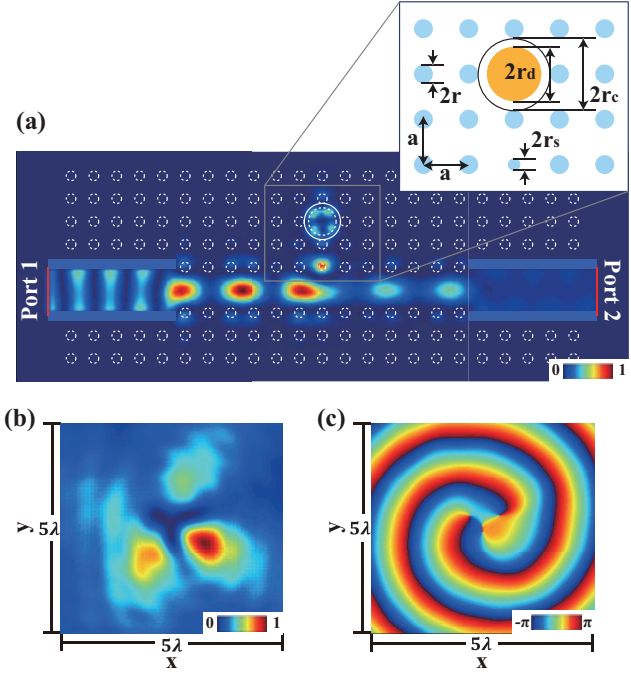}
	\caption{The generation of OAM of order $2$ by the proposed 3D PC. (a) The geometry and corresponding electric-field distribution: the guided wave from port 1 is coupled to the defect modes that radiate out through a circular opening. (b) The intensity and (c) phase distributions of $E_z$ at a transverse plane $30$~mm above the structure. The geometric parameters are as follows: $a=12$ mm, $r=2.5$ mm, $r_d=7.2$ mm, $r_s=1.65$ mm, $r_c=9.5$ mm, and height $h=9.2$ mm. The operating frequency is $8.8$~GHz.}
	\label{quad_sim_3d}
\end{figure}

To make the localized field radiate out of the 3D PC structure, a circular opening of radius $r_c$ is made right above the defect rod, on the top plate of the parallel-plate waveguide, as presented in Fig.~\ref{quad_sim_3d}(a). To excite the two quadrupole modes simultaneously with a phase difference of $90^\circ$, the operation frequency is set to be distant from the two eigenfrequencies. The structure is optimized to generate a purely radiative OAM state in air. The electric field on the $xy$ plane at the operating frequency of $8.8$~GHz is shown in Fig.~\ref{quad_sim_3d}(a). Figures~\ref{quad_sim_3d}(b,c) depict the $z$ component of the radiated electric field at a transverse plane $30$~mm above the structure. The asymmetric quadrupole pattern in Fig.~\ref{quad_sim_3d}(b) results from the asymmetric excitation from the small scatterer. The phase distribution demonstrates an OAM of order 2. We consider the conversion efficiency as the ratio of the power radiated from the circular opening to the incident power at Port 1, and it is calculated to be $19.4\%$.

\begin{figure}[htbp]
	\centering
	\includegraphics[width=\columnwidth]{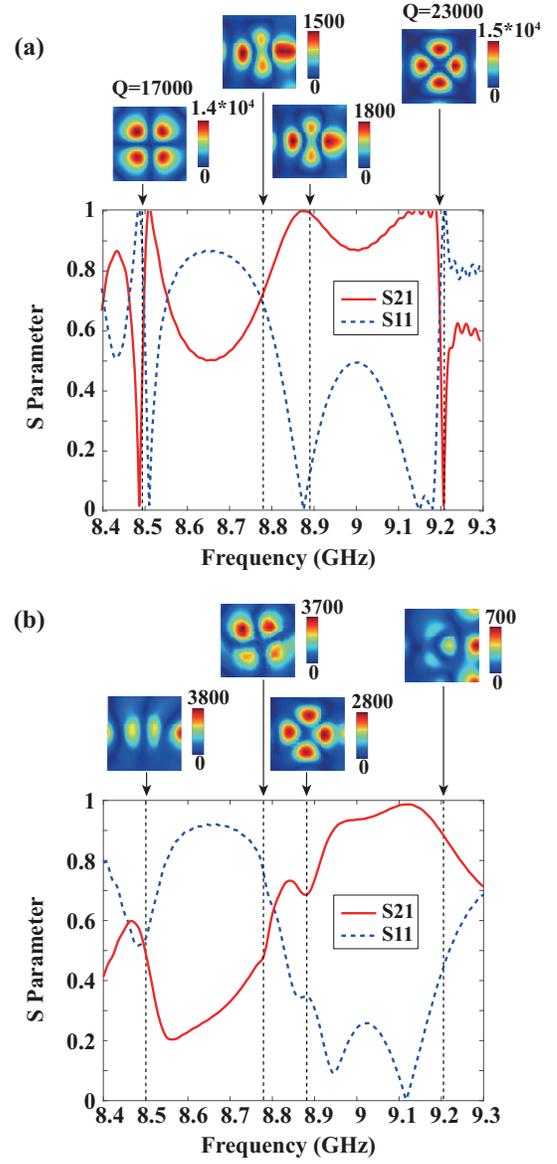}
	\caption{The S parameter and quadrupole modes about the defect at featured frequencies in the (a) 2D PC and (b) 3D PC. $E_z$ is plotted for each mode and the units are V/m. The values are the simulated values under the same input power.}
	\label{quadrupole_Q}
\end{figure}

\begin{figure}[htbp]
	\centering
	\includegraphics[width=\columnwidth]{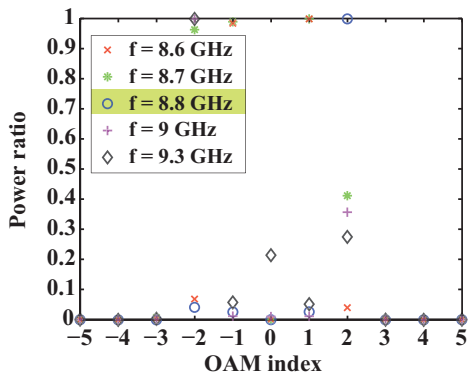}
	\caption{The projection of the complex field, which is above the 3D PC operated in the quadrupole state, along the azimuthal direction onto the orthonormal basis of $e^{il\phi}$.}
	\label{quadrupole_freq}
\end{figure}

The reason for the high outcoupling efficiency is studied. Regarding the 2D PC, the defect with the surrounding rods forms a good cavity and the quality (Q) factors are tens of thousands, as shown in Fig.~\ref{quadrupole_Q}(a). The corresponding eigenfrequencies are well separated from each other. We see a noticeable decrease in the field intensity around the defect when the frequency is distant from the eigenfrequencies. However, for the 3D PC, due to the introduction of the circular opening and the radiation loss through it, the two eigenfrequencies shift and become closer to each other. The quadrupole-xy and quadrupole-diag modes are located at $8.78$ and $8.89$~GHz, respectively, as shown in Fig.~\ref{quadrupole_Q}(b). Due to the radiation loss, the $3$ dB bandwidth at the two eigenfrequencies cannot be identified, so exact values of the Q factors are not calculated. By examining the field amplitudes around the defect, it can be seen that the radiation loss weakens the confinement of the defect mode, indicating the effective outcoupling of the two quadrupole modes at the intermediate operating frequency. In view of the frequency shift, below $8.78$~GHz, we can observe a dipole component. For example, as shown in Fig.~\ref{quadrupole_freq}, when $f=8.7$~GHz, the dipole components are much stronger than the quadrupole components; and when the frequency keeps decreasing to $8.6$ GHz, only dipole states remain. On the other hand, when the frequency goes higher than $8.89$~GHz, the monopole-2 mode is excited. In Fig.~\ref{quadrupole_freq}, a remarkable monopole component can be observed at $9.3$~GHz. 

Additionally, based on our analyses and simulations, two main mechanisms will affect the weights and phase of the two quadrupole modes. The first one is the reflection and refraction at the top air-dielectric interface. Therefore, the height of the rods $h$, the size of the aperture $r_c$, and the thickness of the metallic plate need to be optimized. Among all the parameters, the influence of the height is significant, because the height directly determines the excited modes at the air-dielectric interface (for details, see the Supplemental Material~\cite{SI}). As for the aperture, we choose it to be slightly larger than the concentration region of the field. If the aperture is too small, the radiated field will be distorted. When the aperture is large enough, the radiated field retains the features of the field around the defect. The influence of the aperture is less significant than that of the height; and an optimal aperture size can be obtained in the simulation. Additionally, a thicker top metallic plate will be detrimental to the radiated OAM wave from the aperture$-$but its thickness does not matter as long as it remains thin enough.  The second mechanism is the scattering and interference of waves at the waveguide-scatterer-defect channel. The influence of feeding network and the size of the small scatterer fall into this mechanism. In the proposed design, we feed the 3D PC structure from one port by using the guided wave inside the line defect. The small scatterer is introduced to enhance the mode conversion between the guided mode and the localized defect mode. It functions as a mode coupler that maximizes the coupling between the feeding channel and the defect, and minimizes the transmitted power to the opposite port. By tuning the size of the scatterer, the coupling efficiency from the channel to the defect varies, so do the weights of the quadrupole modes. There is a tradeoff between the high efficiency and purity of the produced OAM wave.

Additionally, the geometric parameters are scaled to shift the operating frequency from the microwave to the optical regime. At optical frequencies, the two metallic plates are replaced by two aluminum plates and the dispersion of aluminum is taken into consideration~\cite{al_dispersion} (for details, see the Supplemental Material~\cite{SI}).

\subsection{OAM generation from dipole defect states}

We make use of other defect modes to generate OAM of different orders. Here, we study the case for the generation of OAM of order $\pm 1$ by superposing the dipole states. Importantly, there is a critical difference between the quadrupole and dipole modes: the dipole modes are doubly degenerate, while the quadrupole modes are splitted. The degenerate dipole modes cannot be exploited to induce a rotational mode carrying OAM, which is demonstrated and explained in the Supplemental Material~\cite{SI}.

\begin{figure}[htbp]
\centering
\includegraphics[width=\columnwidth]{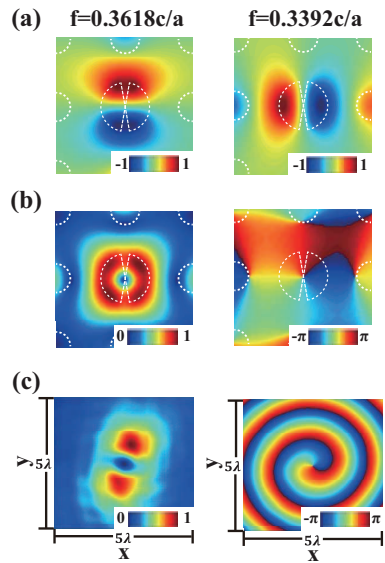}
\caption{The generation of OAM of order $1$. (a) The real part of electric field of the two nondegenerate dipole modes by numerical simulations. Two identical sectors with an angle of $20^\circ$ are cut out from the defect rod ($r_{d}=0.4a$). (b) The amplitude and phase patterns of the superposed electric field. The proper weights and phase in Eq.~(1) are chosen to make the $l=-1$ mode dominant. (c) The full-wave simulated intensity and phase distributions of $E_z$ at a transverse plane $30$~mm above the 3D PC. The geometry of the 3D PC is the same as is shown in Fig.~\ref{quad_sim_3d}(a), except for the cut in the defect rod. The geometric parameters are as follows: $a=12$ mm, $r=2.5$ mm, $r_d=4.8$ mm, $r_s=1.3$ mm, $r_c=6$ mm and $h=7$ mm. The operating frequency is $9.7$~GHz.}
\label{dipole}
\end{figure}

To create an analog to the quadrupole case, we break the $C_4$ rotational symmetry of the defect by introducing two sector cuts. Consequently, the degenerate dipole modes split into two dipole modes with different eigenfrequencies, as shown in Fig.~\ref{dipole}(a). Based on the variational principle~\cite{pcbook1}, concentration of field into the higher-permittivity material decreases the mode frequency, which is consistent with the numerical results. Similarly, by superposing the two modes with selected weights and phase values, an electric field with the spatial phase dependence of $e^{\pm i\phi}$ can be obtained. Figure~\ref{dipole}(b) shows one of the cases with $l=-1$.

To effectively excite the two dipole modes, the defect is rotated $45^\circ$ (for details, see the Supplemental Material~\cite{SI}). In Fig.~\ref{dipole}(c), we show the $z$ component of the radiated electric field. An EM wave with an OAM of order 1 is generated. The conversion efficiency is calculated to be $24.9\%$.

\begin{figure}[htbp]
	\centering
	\includegraphics[width=\columnwidth]{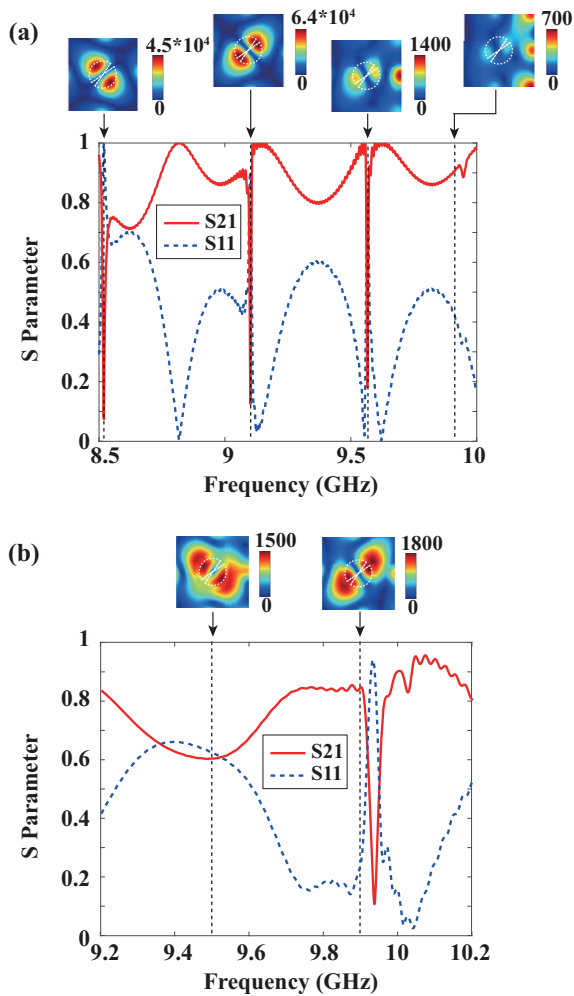}
	\caption{The S parameter and dipole modes about the defect at feature frequencies in the (a) 2D PC and (b) 3D PC. $E_z$ is plotted for each mode and the units are V/m. The values are the simulated values under the same input power.}
    \label{dipole_Q}
\end{figure}

It should be emphasized that the operating frequency of $9.7$~GHz in Fig.~\ref{dipole}(c) is higher than the two eigenfrequencies in the 2D PC, which are around $8.5$ and $9.1$~GHz [Fig.~\ref{dipole_Q}(a)]. In the 2D PC, when the operating frequency goes higher than $9.1$~GHz, the dipole mode becomes weakened and distorted. When the circular opening is inserted, the two eigenfrequencies shift, which is similar to the quadrupole case. In the 3D PC, the two dipole modes are shifted to $9.5$ and $9.9$ GHz [Fig.~\ref{dipole_Q}(b)] with less confinement around the defect, which indicates the effective outcoupling of the two dipole modes at $9.7$~GHz.

\subsection{Discussion}

To generate OAM of order -1 and -2, we mirror the whole structure, including the feeding ports about $yz$ plane. Based on this transformation rule ($x\to -x,\,\,y\to y$), the phase of one vibrational mode is flipped and that of the other remains unchanged. As a result, the excited rotational mode will reverse its helicity. Specifically, for the quadrupole structure, by feeding from the receiving port (Port 2) in Fig.~\ref{quad_sim_3d}(a), the radiated wave will carry OAM of order $-2$. Our theory is verified by full-wave simulations, as presented in the Supplemental Material~\cite{SI}.

\section{Experiments}

\begin{figure}[htbp]
\centering
\includegraphics[width=0.9\columnwidth]{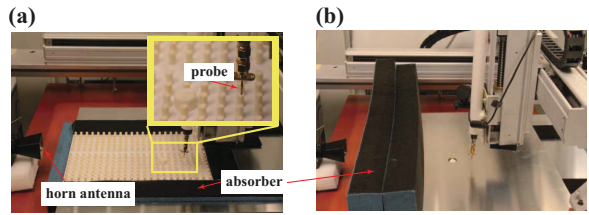}
\caption{The experimental setup. (a) The measurement platform without the top aluminum plate. (b) The complete view of the measurement platform.}
\label{experiment}
\end{figure}

The generation of OAM by the proposed 3D PCs is verified by experiments. The implementation of the PCs in microwave regime adopts the approach in Refs.~\cite{CTChan1,CTChan2}. The PCs are made of alumina rods with $\epsilon_{r}=8.5$ (for details, see the Supplemental Material~\cite{SI}). They are assembled between two flat aluminum plates, the top one of which has a cutting hole. The whole structure is surrounded by absorbers to avoid unwanted scattering. The experimental setup is shown in Fig.~\ref{experiment}. A standard linearly polarized horn antenna operating from $6.57$ to $9.99$~GHz is connected to the transmitting port of a vector network analyzer (Agilent E5071C). The center of the horn antenna is aligned with the central line of the line defect in the assembled PC structure, in order to feed more power into the channel. A probe made from a $50 \Omega $ SMA connector is attached to the receiving port of the vector network analyzer. The probe is vertically polarized and fixed onto a near-field scanning platform to measure the vertically polarized field component ($E_z$) at a transverse plane. Additional absorbers are placed near the horn antenna on the top of the top alumina plate to prevent direct transmission from the horn antenna to the receiving probe [(Fig.~\ref{experiment}(b)]. During the measurement, the time-domain gate technique is applied to prevent unwanted multiple reflections. By controlling the scanner, the field at the transverse plane is sampled every $10$~mm. The distance between the probe and the top aluminum plate is $25$~mm. The ratio of the received voltage to the transmitted voltage provides the magnitude and phase of the field at the probe position.

\begin{figure}[htbp]
\centering
\includegraphics[width=0.9\columnwidth]{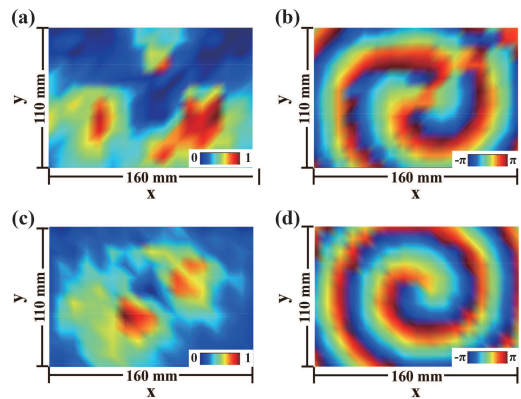}
\caption{The intensity and phase distributions of the measured $E_z$ at a transverse plane $25$~mm above the top aluminum plate of the 3D PCs. (a) The intensity and (b) phase distributions of the quadrupole case: the frequency is $8.75$~GHz. (c) The intensity and (b) phase distributions of the dipole case: the frequency is $9.55$~GHz.}
\label{exp_result}
\end{figure}

The field over a 160 mm x 110 mm area is measured. During the measurement, we sweep the frequency of the vector network analyzer and the frequency corresponding to the optimal field pattern is chosen for the plots in Fig.~\ref{exp_result}. The working frequencies for both the dipole and quadrupole cases are found to be slightly lower than the simulated ones. This divergence is reasonable, in view of fabrication errors, assembling errors, permittivity deviation, etc. Overall, the OAM states are verified by experiments and the results are in good agreement with the numerical simulations.

\section{CONCLUSION}
In conclusion, we show that by superposing two vibrational defect states, a rotational defect state carrying OAM can be produced. The proposed 3D PCs couple the guided wave in a line defect to the superposed localized state in a point defect. With a circular opening on the top, the localized defect mode leaks out. One small scatterer between the channel and the defect is employed to improve the coupling efficiency. The lowest four OAM states, i.e., $\pm1$ and $\pm2$, are generated and verified by both simulations and experiments. The working mechanism of the 3D PC structures provides a route for OAM generation. Higher-order OAM waves may be produced by continuously increasing the radius or permittivity of the defect so that higher-order localized modes will appear and be manipulated~\cite{tune_defect}.

\begin{acknowledgments}
This work was supported in part by the Research Grants Council of Hong Kong (GRF 716713, GRF 17207114, and GRF 17210815), NSFC 61271158, Hong Kong UGC AoE/P-04/08, AOARD FA2386-17-1-0010, Hong Kong ITP/045/14LP, Hundred Talents Program of Zhejiang University (No. 188020*194231701/208), and HKU Seed Fund 201711159228.
\end{acknowledgments}


\end{document}